\documentclass[preprint,12pt]{elsarticle}
\usepackage{amsmath}
\usepackage{amssymb}
\usepackage[pdftex,colorlinks]{hyperref}
\usepackage{multirow}
\usepackage{enumerate}
\usepackage{cases}
\usepackage{graphicx}
\newtheorem{Protocol}{Protocol}

\usepackage{floatrow}
\newfloatcommand{capbtabbox}{table}[][\FBwidth]

\usepackage{caption}
\usepackage{subfigure}

\usepackage{bbding}

\newtheorem{Definition}{Definition}

\begin{document}

\begin{frontmatter}

\title{Quantum Bit Commitment Protocol Based on Counterfactual Quantum Cryptography}


\author{Ya-Qi Song $^{1,2,3}$}
\author{Li Yang$^{1,2}$\corref{1}}
\cortext[1]{Corresponding author. Email: yangli@iie.ac.cn}
\address{1.State Key Laboratory of Information Security, Institute of Information Engineering, Chinese Academy of Sciences, Beijing 100093, China\\
2.Data Assurance and Communication Security Research Center,Chinese Academy of Sciences, Beijing 100093, China\\
3.School of Cyber Security, University of Chinese Academy of Sciences, Beijing 100049, China}

\begin{abstract}
We present a new quantum bit commitment (QBC) protocol based on counterfactual quantum cryptography. We analyze the security of this protocol, find that it can resist the attack presented by QBC's no-go theorem. Our protocol is simple, and probably gives a new way of constructing QBC protocol.

\end{abstract}


\end{frontmatter}


\section{Introduction}
\label{intro}
The bit commitment (BC) scheme is a basic primitive of modern cryptography. The concept of bit commitment was first proposed by Blum in \cite{Blum82}. It plays a crucial role in constructions of multi-party protocols, such as zero-knowledge proof scheme, verified secret shared scheme, and so on.

BC scheme includes two phases, i.e. the commit phase and the opening phase. In the commit phase, Alice chooses a commit bit $b$ and sends a piece of evidence to Bob. In the opening phase, Alice unveils the value of $b$ and Bob checks it. A commit bit commitment scheme has the following properties. (i) Correctness. If Alice and Bob execute the scheme honestly, Bob obtains the correct commit bit $b$ in the opening phase. (ii) Concealing. Bob cannot know the commit bit $b$ before the opening phase. (iii) Binding. Alice cannot change the commit bit after the commit phase. A BC scheme is unconditionally secure if there is no computational assumption on attacker's ability and it satisfies the properties of concealing and binding.

The first quantum bit commitment (QBC) scheme was proposed in 1984\cite{BB84} and it pointed that the binding security of the scheme can be attack by sending entangled states. In 1993, a well-known QBC scheme was presented \cite{BCJL}, which is usually referred to as BCJL scheme and was once believed as a provably secure scheme. Unfortunately, Mayers found that the BCJL scheme was insecure \cite{Mayers96}. Later, Mayers, Lo and Chau separately present no-go theorem and prove that the unconditional secure QBC protocol is impossible \cite{Mayers97,Lo97,BCMno-go97}.

Although the correctness of no-go theorem is undoubtedly, the framework of the theorem may not cover all the types of QBC protocols. People try to construct QBC protocols to resist the attack presented by no-go theorem  under some security conditions. Using the different agents, the relativistic QBC protocols are proposed by Kent \cite{Kent99,Kent05,Kent12}. The protocol \cite{Kent12} has been implemented by different groups \cite{shiyan1,shiyan2}. There are several QBC protocols proposed based on physical hypothesis, such as bounded-quantum-storage model \cite{DF05,DD07}, noisy-storage model \cite{WS08,NJ12,KW12} and technological limitations on non-demolition measurements \cite{Vaidman12}. In \cite{SY15}, Song and Yang construct  practical quantum one of two oblivious transfer and QBC protocols based on \cite{Yang13} with physical security.

GP He finds a new way to construct QBC scheme \cite{He11,He14} based on orthogonal quantum key distribution (QKD) scheme \cite{GV,KI}. In his QBC scheme, Alice plays the role of the two parties in the orthogonal QKD scheme, while Bob plays the eavesdropper in the QKD scheme. The orthogonal states of GP He's schemes are the key point to resist the attack presented by no-go theorem. Inspired by his construction, we propose a QBC protocol based on counterfactual quantum cryptography \cite{CQKD}. In our protocol, although Bob plays the role of the eavesdropper in counterfactual QKD, both Alice and Bob are senders and receivers. Even though our construction is inspired by GP He's scheme, the reason of our protocol evading no-go theorem is different from his. The main reason is that Alice cannot distinguish which qubits arrive at her site and which qubits she should apply the local unitary transformation on, which probably gives a new way of constructing QBC protocol.

The remainder of the paper is organized as follows. The next section is the preliminaries of counterfactual QKD and its security. Section 3 gives a universal framework for QBC protocol. We propose our QBC protocol in Section 4. Security analysis on probability and choosing the appropriate security parameters are in Section 5 and Section 6, respectively. In Section 7, we discuss the reason that our protocol evades no-go theorem. The possible development and the main ideas of our protocol are summarized in the discussion and conclusion sections.

\section{Preliminaries}
\subsection{Counterfactual quantum cryptography}

\begin{figure}
\centering
\includegraphics[width=0.9\textwidth]{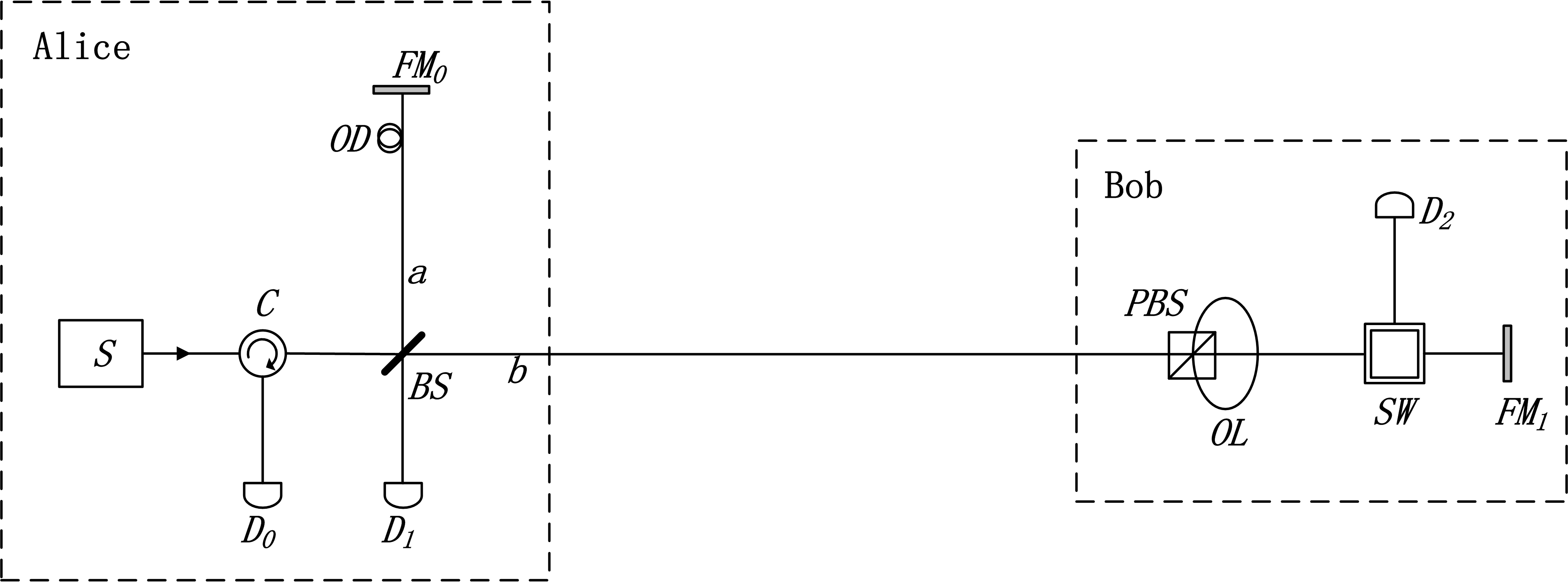}
\caption{The architecture of the N09 QKD protocol. The setup is a modification based on Michelson-type interferometer. The single photon source $S$ emits a optical pulse containing only one photon. Then the pulse is transmitted through the optical circulator $C$ and split into two pulses by the beam splitter $BS$. The two light paths $a$ and $b$ are the arms of the Michelson-type interferometer, and the length of the path $a$ is adjusted by an optical delay $OD$. The pulse transmitted through path $a$ is reflected by the Faraday mirror $FM_0$ and back to $BS$. The pulse transmitted through path $b$ travels to Bob's site. If the pulse is horizontally polarized, it pass through the polarizing beam splitter $PBS$, or it is reflected by $PBS$ and pass through the optical loop $OL$. The arriving time to the optical switch $SW$ of the different polarized pulses is different. Only the $SW$ controlled in the correct time, the pulse will reach the detector $D_2$. Otherwise, the pulse will be reflected by $FM_1$ and return back to Alice's site. The back pulse from path $b$ and the pulse from path $a$ are combined at the $BS$ and interfered to lead the detector $D_0$ to click. }
\label{fig:QKD}       
\end{figure}

\cite{CQKD} proposed a special QKD protocol (N09), in which the particle carrying secret information is not transmitted through the quantum channel. Fig.~\ref{fig:QKD} shows the architecture of the QKD protocol.
In the QKD protocol, Alice randomly encodes horizontal-polarized state $|H\rangle$ as the bit value $"0"$ or vertical-polarized state $|V\rangle$ as the bit value $"1"$ and sends the state by the single photon source $S$. When Bob's bit value is the same as Alice's, the optical switch $SW$ controlled in the correct time. In this case, the interference is destroyed and there are three occasions for the single photon. Suppose the reflectivity and transmissivity of the $BS$ are $R$ and $T$, where $R+T=1$. (i) Detector $D_0$ clicks with the probability of $R^2$. The photon travels via path $a$ and then is reflected by the $BS$ again. (ii) Detector $D_1$ clicks with the probability of $RT$. The photon travels via path $a$ and then pass though the $BS$. (iii) Detector $D_2$ clicks with the probability of $T$. The photon travels via path $b$ and is controlled by the $SW$ to reach the detector $D_2$. When Bob's bit value is different from Alice's, the setup is a Michelson-type interferometer and the detector $D_0$ clicks. Alice and Bob only remain the bit in the event that the detector $D_1$ clicks alone to be the shared keys. The other events are used for eavesdropping detection.
\subsection{The security of counterfactual quantum cryptography}
\label{sec:1}
The key point of the N09 protocol is that the final shared keys is made up of the pulses only travel via the path $a$ not the path $b$. Actually the secret keys are not transmitted through the quantum channel. In \cite{securityQKD10}, Yin et al. proposed an entanglement distillation protocol equivalent to the N09 protocol. Then give a strict security proof assuming that the perfect single photon source is applied and Trojan-horse attack can be detected. In 2012, Zhang et al. give a more intuitive security proof against the general intercept-resend attacks \cite{securityQKD12}. These show that the N09 protocol is secure in ideal conditions with infinite recourses.  An attack for the protocol that the devices used by Alice and Bob are perfect but the length of the generated key is finite was presented in \cite{attack}.

\begin{figure}
\centering
\includegraphics[width=0.9\textwidth]{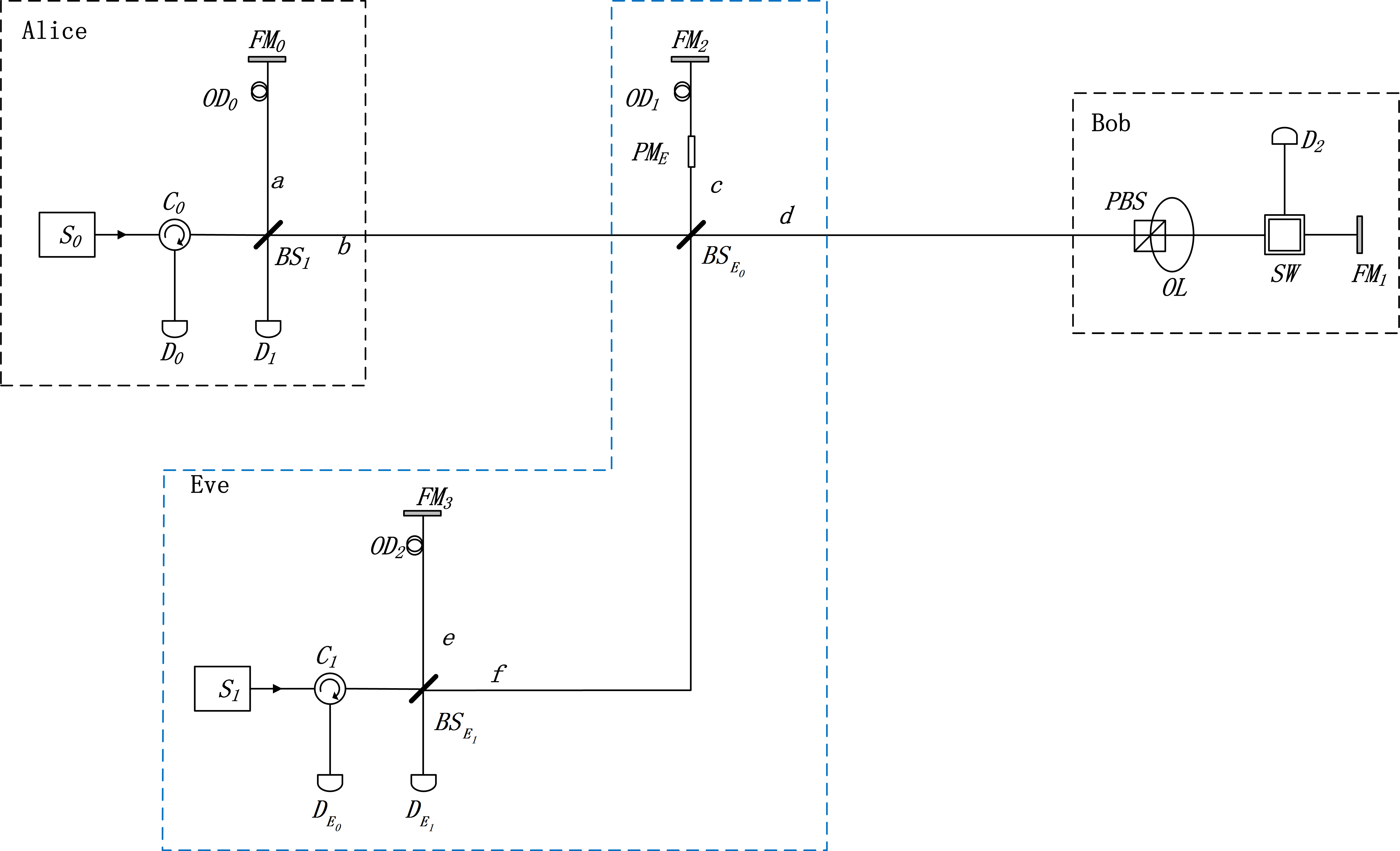}
\caption{The eavesdropping strategy of N09 protocol. Based on the original architecture of N09 protocol, Eve prepares the same apparatus as Alice's and places them by the router module. The router module contains the beam splitter $BS_{E_1}$, the phase modulator $PM_E$, optical delay lines $OD_1$ and Faraday mirror $FM_2$. }
\label{fig:attack}       
\end{figure}

The eavesdropping strategy of \cite{attack} is shown in Fig.~\ref{fig:attack}. The eavesdropping scheme is outlined below. (i) When Alice sends a photon to Bob, Eve also sends a random polarized photon at the same time. (ii) Eve records which detectors click. (iii) After Alice and Bob publish their detection results, Eve can extract secret information when both of the detectors $D_1$ and $D_{E_1}$ click.

We first analyze the photon sent by Alice ($S_0$). The photon is transmitted through the optical circulator $C_0$ and split by he beam splitter $BS_1$ into two wave packets. The wave packet through path $a$ stays in Alice's site while the wave packet through path $b$ passes through the beam splitter $BS_{E_0}$ and evolves into two wave packets through paths $c$ and $d$. The wave packet through path $c$ may travel back to Alice's site or Eve's site. The wave packet through path $d$ travels to Bob's site. If Bob's bit is different from  Alice's, the wave packet through path $d$ is reflected by the Faraday mirror $FM_1$ and back to the beam splitter $BE_{E_0}$. The optical delay $OD_1$ and the phase modulator $PM_E$ are used to adjust the wave packet through path $c$ to interfere with the wave packet through path $d$. Then it travels to the beam splitter $BS_1$ and interfere with the wave packet through path $a$ and makes the detector $D_0$ click. If Bob's bit is the same as Alice's, the interference is destroyed. The detectors $D_0$, $D_1$ and $D_2$ are all possible to click. The condition of the photon sent by Eve is similar to the analysis above. After Alice and Bob announce the detection results, Eve knows that the polarization of the information qubit is the same as his when detectors $D_1$ and $D_{E_1}$. The addition of another detector leads to multiple photon detection and the detection probability of $D_0$, $D_1$ and $D_2$ may be changed. However, by analysis the probability of all the detection, the error rate introduced by the attack tend to zero simultaneously as $r_{E_0}\rightarrow 0$, where $r_{E_0}$ is the reflectivity of the beam splitter $BS _{E_0}$. To ensure the correlation between Bob and Eve, $r_{E_0}$ never reach the value $0$. Although this attack cannot work in the ideal N09 protocol with infinite resources, the error rate can be due to the imperfect devices in the practical with finite resources.

\section{A framework for QBC}
QBC is a two-party cryptographic protocol. In the commit phase, one party Alice commits to the other party Bob to a bit $b$ by sending a piece of evidence. In the opening phase, Alice announces the value of $b$ and Bob verifies whether it is indeed the commit bit. We define a kind of quantum channel to construct QBC protocol.
\begin{Definition}
  \emph{\textbf{QBC Channel}}\\
  In commit phase, Alice sends quantum states as a piece of evidence of commit bit to Bob via a channel, for each qubit if
  \begin{enumerate}
    \item The probability of Bob knowing for sure what Alice sends is denoted $P_B$, $0<P_B<1$;
    \item The probability that Alice confirms that Bob obtains her state is denoted $P_A$, $0\leq P_A<P_B$.
  \end{enumerate}
  Then, the channel is named a QBC channel.
\end{Definition}
 $P_B>0$ means that Bob has a piece of evidence. Because of $P_B<1$, Bob cannot know all of the qubits correctly. By choosing appropriate security parameters $n$, the protocol can satisfy the concealing security.  If Alice tries to alter one bit in the opening phase, her best choice is to change the bit she cannot distinguish whether Bob knows with a probability of $1-P_A$. In fact, there are around $(1-P_B)n$ qubits Bob cannot judge. If $P_A=P_B$, Alice can accurately alters the bit in part that Bob really does not know without detection. If $P_A<P_B$, the range of bits Alice alters $1-P_A$ is larger than  Bob cannot distinguish, and her attack may be caught. Therefore, the conditions $0\leq P_A<P_B<1$ is the necessary condition of the binding security.

\section{QBC based on N09}
We use the N09 protocol to construct our QBC protocol. As the security of the N09 protocol has been proved and the key point of the attack for finite resources introduced is to control the reflectivity of the beam splitter $BS _{E_0}$ (see Sect.~\ref{sec:1}). However, $r_{E_0}\rightarrow 0$ can only protect Eve from detection. She cannot obtain all the information by the devices in Fig.~\ref{fig:attack} with finite detectors.

-

\begin{figure}
\centering
\includegraphics[height=5.2cm]{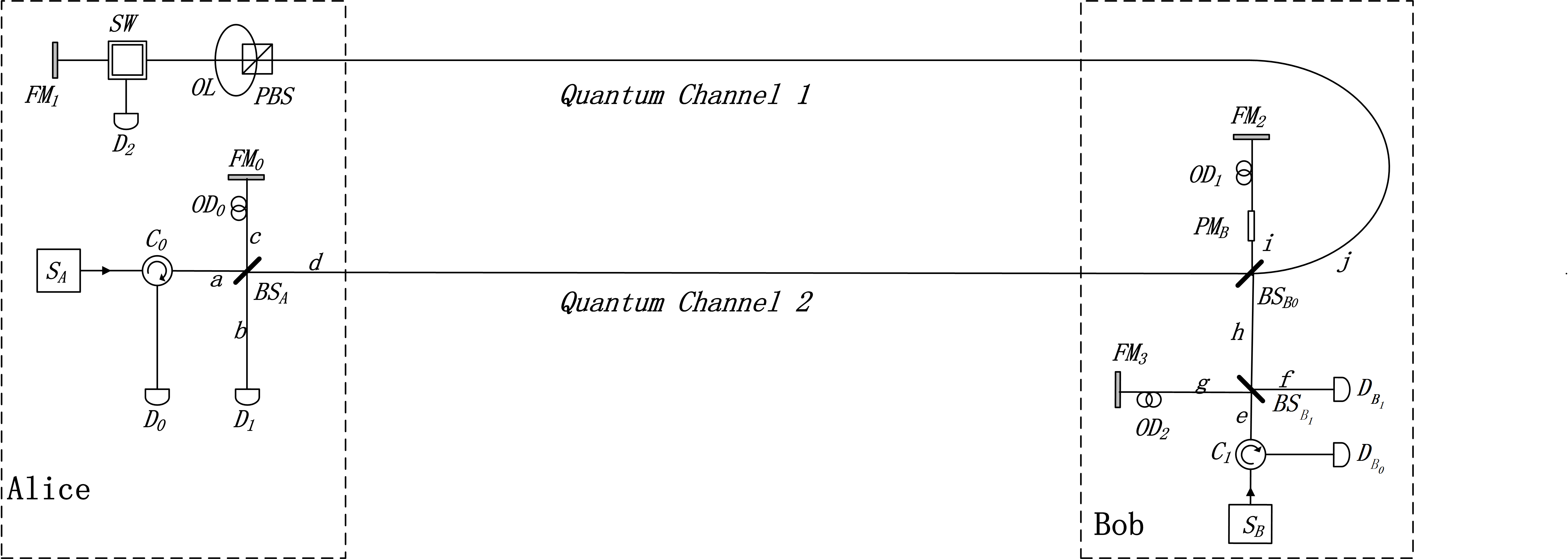}
\caption{The architecture of our QBC protocol. Alice's apparatus are the same as the sender and receiver of N09 protocol while Bob's apparatus are the same as the eavesdropping of \cite{attack}. As the two parties of the QBC protocol, the distance between them is $L$. The length of the optical delay line $OD_0$ is around $2L$ and the lengths of the optical delay line $OD_1$ and $OD_2$ are around $L$.}
\label{fig:QBC}       
\end{figure}

The party Alice in the QBC protocol plays the role of the both parties in the N09 protocol. And the behavior of the party Bob in the QBC protocol is the same as that of Eve in the N09 protocol. The information Bob obtains is treated as the piece of the evidence in the commit phase, which cannot unveil the value of $b$ for Bob and ensures Alice not to change it. Fig.~\ref{fig:QBC}  shows the architecture of our QBC protocol.

\begin{Protocol}
\emph{\textbf{Counterfactual quantum bit commitment}}
~\\
Commit Phase:
\begin{enumerate}
\item Alice and Bob set up devices according to Fig.~\ref{fig:QBC}, where the beam splitters $BS_A$, $BS_{B_0}$ and $BS_{B_1}$ are half transparent and half reflecting mirrors. The detection devices in the Fig.~\ref{fig:QBC} are simplified graphics which can detect the number and the polarization of the photons. They agree on two security parameters $m$ and $n$.
\item Alice chooses a random bit $b \in{\{0,1\}}$ as her commit bit. Then she generates $m$ random bit strings according to the value of $b$. Each sequence consists $n$ bits, which can be represented as $a^{(i)}\equiv ({a_1^{(i)}}{a_2^{(i)}}...{a_n^{(i)}})\in{\{0,1\}}^{n}$, $i=1, 2,..., m$. The sequences satisfy ${a_1^{(i)}}\oplus{a_2^{(i)}}\oplus...\oplus{a_n^{(i)}}=b$.
\item Bob generates $m$ bit strings randomly and uniformly with the length of $n$. Each sequence is represented as $b^{(i)}\equiv ({b_1^{(i)}}{b_2^{(i)}}...{b_n^{(i)}})\in{\{0,1\}}^{n}$.
\item Alice and Bob decide on a series of time instants $t_1^{(i)}, t_2^{(i)}, ..., t_n^{(i)}$ and $\bigtriangleup t$, where $\bigtriangleup t$ is the time a photon spends from the beam splitter $BS_{B_1}$ to $BS_{B_0}$. Alice sends $|\Psi_{a_j^{(i)}}\rangle$ at the time $t_j^{(i)}$ while Bob sends $|\Psi_{b_j^{(i)}}\rangle$ at the time $t_j^{(i)}+\bigtriangleup t$ by their own single photon sources, respectively. $|\Psi_{0}\rangle=|H\rangle$ and $|\Psi_{1}\rangle=|V\rangle$ represent the horizontal-polarized state and the vertical-polarized state, respectively.
\item Alice and Bob record the time and response of their detectors.
\end{enumerate}
~\\Opening Phase:
\begin{enumerate}
\item Alice reveals the commit bit $b$, the $m$ sequences $({a_1^{(i)}}{a_2^{(i)}}...{a_n^{(i)}})$ and the response of her three detectors to Bob.
\item Bob verifies whether ${a_1^{(i)}}\oplus{a_2^{(i)}}\oplus...\oplus{a_n^{(i)}}=b$, and the response of all the detectors agree with the state $|\Psi_{a_j^{(i)}}\rangle$. If the consistency holds, he admits Alice's commit value as $b$.
\end{enumerate}
\end{Protocol}

\section{Security analysis}
\subsection{Basic ideas}
The response of the detectors is according to the consistency of the bits chosen by Alice and Bob, which keeps secret in commit phase.
By the response of detectors $D_{B_0}$ and $D_{B_1}$, Bob knows Alice's bit for sure with a probability of $P_B$. Alice confirms that Bob knows her bit at the ratio of $P_A$ according to the response of her detectors $D_0$, $D_1$ and $D_2$. If Alice tries to attack the binding of the protocol, she has to alter odd bits for each sequence in the opening phase. In each sequence, she can distinguish that around $P_An$ bits are confirmed by Bob. Alice's optimal strategy is to alter one bit in the range of the other $(1-P_A)n$ bits. Among the $(1-P_A)n$ bits, only $(1-P_B)n$ bits are not known by Bob. Therefore, the probability that Alice alters one bit without detection is
\begin{equation}
  P(Aalter)=\frac{1-P_B}{1-P_A}.
\end{equation}
Then in the QBC scheme, the probability of changing the commit bit without detection is $P(Aatler)^m$. When $P(Aalter)<1$, the probability of breaking the binding security can be arbitrarily close to zero with the increase of $m$. Given a binding security threshold $\alpha$, i.e. $P(Aatler)^m<\alpha$, the range of $m$ is
\begin{equation}\label{threshold of m}
  m>\frac{\log\alpha}{\log P(Aalter)}
\end{equation}

For a sequence of qubits, Bob makes sure the commit value with a probability of $P_B^n$. Given $m$ qubit strings, the probability that Bob has no idea about the commit value is $(1-P_B^n)^m$.  Define $\varepsilon$ as the probability that Bob ascertains the commit value,
\begin{equation}
  \varepsilon\equiv1-\left(1-P_B^n\right)^m.
\end{equation}
If Bob does not confirm the commit value from the protocol, he just guess with a probability of $1/2$. Therefore, the probability that Bob obtians the right commit value is
\begin{equation}
  P(Bknows)=\varepsilon+\frac{1-\varepsilon}{2}=\frac{1}{2}+\frac{\varepsilon}{2}.
\end{equation}
Then the advantage of Bob breaking the concealing security is
\begin{equation}\label{pright}
  \left|P(Bknows)-\frac{1}{2}\right|=\frac{\varepsilon}{2}=\frac{1}{2}-\frac{(1-P_B^n)^m}{2}
\end{equation}
When $P_B^n$ is small, Eq.~(\ref{pright}) approximates to $mP_B^n/2$, which can be arbitrarily close to zero with appropriate security parameters $m$ and $n$. Given a concealing security threshold $\beta$, to satisfy $\left|P(Bknows)-\frac{1}{2}\right|<\beta$ the accurate range of $n$ is
\begin{equation}\label{threshold of n}
  n>\frac{\log[1-(1-2\beta)^{1/m}]}{\log P_B}.
\end{equation}

Based on the analysis above, if the quantum channel is a QBC channel in Definition 1, which satisfies $0\leq P_A<P_B<1$, we can choose appropriate security parameters to make the protocol secure. Then we analyze the response of each detector and the relations between the response and $P_A$, $P_B$, calculate the values of $P_A$, $P_B$ and give the appropriate security parameters.

\subsection{The response of each detector}
In our protocol, the detector $D_2$ is in Alice'site and Alice controls the optical switch $SW$ according to the states sent by herself. Therefore, the photon sent by Alice cannot interfere. All the situations can be classified into two categories, one is $a_j^{(i)}\neq b_j^{(i)}$, the other is $a_j^{(i)}= b_j^{(i)}$.
\begin{enumerate}
  \item When the bits chosen by Alice and Bob are different, i.e.    $a_j^{(i)}\neq b_j^{(i)}$, the photon sent by Bob's source $S_B$ must interfere and return to detector $D_{B_0}$. There are different optical paths to make the detectors click.
  \begin{enumerate}
  \item Detector $D_0$ clicks. There are two possible situations. One is that the photon sent by the source $S_A$ is reflected by the beam splitter $BS_A$, travels through the path $c$, then returns back to the beam splitter $BS_A$ and detector $D_0$ responses, i.e. $S_A \rightarrow C_0 \rightarrow BS_A \rightarrow FM_0  \rightarrow BS_A  \rightarrow C_0 \rightarrow D_0$. For the other path, the photon sent by the source $S_A$ travel through the beam splitter $BS_A$ and is reflected by the beam splitter $BS_{B_0}$, then reflected by the Faraday mirror $FM_2$ back to the path $d$ and arrive to detector $D_0$, i.e. $S_A \rightarrow C_0 \rightarrow BS_A \rightarrow BS_{B_0} \rightarrow FM_2  \rightarrow BS_{B_0} \rightarrow BS_A \rightarrow C_0 \rightarrow D_0$.
  \item Detector $D_1$ clicks. The two paths are similar to those lead $D_0$ click. The difference is that when the photon sent by the source $S_A$ returns to the beam splitter $BS_A$, it goes via the other path to detector $D_1$ instead to $D_0$, i.e. $S_A \rightarrow C_0 \rightarrow BS_A \rightarrow FM_0  \rightarrow BS_A  \rightarrow D_1$ and $S_A \rightarrow C_0 \rightarrow BS_A \rightarrow BS_{B_0} \rightarrow FM_2  \rightarrow BS_{B_0} \rightarrow BS_A \rightarrow D_1$.
  \item Detector $D_2$ clicks. There is only one possible path for this detection. The photon sent by the source $S_A$ travels through the beam splitter $BS_A$ and $BS_{B_0}$ via the path $j$ reach detector $D_2$, i.e. $S_A \rightarrow C_0 \rightarrow BS_A \rightarrow BS_{B_0}  \rightarrow PBS \rightarrow D_2$.
  \item Detector $D_{B_0}$ clicks. There are two possible situations. The photon sent by Bob is detected at detector $D_{B_0}$, i.e. $S_B \rightarrow C_1 \rightarrow BS_{B_1} \rightarrow interference \rightarrow C_1 \rightarrow D_{B_0} $. The photon sent by Alice could go to detector $D_{B_0}$ via the paths $a$, $d$, $i$, $h$ and through the beam splitter $BS_{B_1}$, i.e. $S_A \rightarrow C_0 \rightarrow BS_A \rightarrow BS_{B_0} \rightarrow FM_2 \rightarrow BS_{B_0} \rightarrow BS_{B_1} \rightarrow C_1 \rightarrow D_{B_0}$.
  \item Detector $D_{B_1}$ clicks. There is the only path for this situation that the photon sent by Alice travels via the paths $a$, $d$, $i$, $h$ and is reflected by the beam splitter $BS_{B_1}$ to detector $D_{B_1}$, i.e. $S_A \rightarrow C_0 \rightarrow BS_A \rightarrow BS_{B_0} \rightarrow FM_2 \rightarrow BS_{B_0} \rightarrow BS_{B_1} \rightarrow D_{B_1}$.
  \end{enumerate}

  \item When the bits chosen by Alice and Bob are the same, i.e. $a_j^{(i)}= b_j^{(i)}$, there is no interference. The paths which make the detectors click are as follows.
      \begin{enumerate}
        \item Detector $D_0$ clicks. There are three different paths to make detector $D_0$ click. Two of them are the same as the situation $a_j^{(i)}=b_j^{(i)}$. And the third path is that the photon sent by the source $S_B$ travels through the paths $e$, $h$, $i$, $d$, then goes through the beam splitter $BS_A$ and arrives at detector $D_0$. They are $S_A \rightarrow C_0 \rightarrow BS_A \rightarrow FM_0  \rightarrow BS_A  \rightarrow C_0 \rightarrow D_0$, $S_A \rightarrow C_0 \rightarrow BS_A \rightarrow BS_{B_0} \rightarrow FM_2  \rightarrow BS_{B_0} \rightarrow BS_A \rightarrow C_0 \rightarrow D_0$, and $S_B \rightarrow C_1 \rightarrow BS_{B_1} \rightarrow BS_{B_0} \rightarrow FM_2 \rightarrow BS_{B_0} \rightarrow BS_A \rightarrow C_0 \rightarrow D_0$.
        \item Detector $D_1$ clicks. The three paths are the similar to those of detector $D_0$. When the photon travels back to the beam splitter $BS_A$, it goes the way to detector $D_1$. They are $S_A \rightarrow C_0 \rightarrow BS_A \rightarrow FM_0  \rightarrow BS_A  \rightarrow D_1$, $S_A \rightarrow C_0 \rightarrow BS_A \rightarrow BS_{B_0} \rightarrow FM_2  \rightarrow BS_{B_0} \rightarrow BS_A \rightarrow D_1$, and $S_B \rightarrow C_1 \rightarrow BS_{B_1} \rightarrow BS_{B_0} \rightarrow FM_2 \rightarrow BS_{B_0} \rightarrow BS_A \rightarrow D_1$.
        \item Detector $D_2$ clicks. Two different paths make the detector response. One is the same as the situation $a_j^{(i)}=b_j^{(i)}$. The other one is that the photon sent by Bob travels via the paths $e$, $h$, $j$ and get to the detector, i.e. $S_B \rightarrow C_1 \rightarrow BS_{B_1} \rightarrow BS_{B_0} \rightarrow PBS \rightarrow D_2$.
        \item Detector $D_{B_0}$ clicks. There are three possible situations. One is that the photon sent by the source $S_A$ travels via the paths $a$, $d$, $i$, $h$, $e$ to the detector, i.e. $S_A \rightarrow C_0 \rightarrow BS_A \rightarrow BS_{B_0} \rightarrow FM_2 \rightarrow BS_{B_0} \rightarrow BS_{B_1} \rightarrow C_1 \rightarrow D_{B_0}$. The second one is that the photon sent by Bob travels via the paths $e$, $h$, $i$, $h$ to the beam splitter $BS_{B_1}$, then leads the detector click, i.e. $S_B \rightarrow C_1 \rightarrow BS_{B_1} \rightarrow BS_{B_0} \rightarrow FM_2 \rightarrow BS_{B_0} \rightarrow BS_{B_1} \rightarrow C_1 \rightarrow D_{B_0}$. The last one is that the photon sent by Bob is reflected by the beam splitter $BS_{B_1}$ and the Faraday mirror $FM_3$, then through the optical circulator $C_1$ to the detector, i.e. $S_B \rightarrow C_1 \rightarrow BS_{B_1} \rightarrow FM_3 \rightarrow BS_{B_1} \rightarrow C_1 \rightarrow D_{B_0}$.
        \item Detector $D_{B_1}$ clicks. The paths are similar to those in the last item. The difference is that when the photon return to the beam splitter $BS_{B_1}$, it goes directly to detector $D_{B_1}$, i.e. $S_A \rightarrow C_0 \rightarrow BS_A \rightarrow BS_{B_0} \rightarrow FM_2 \rightarrow BS_{B_0} \rightarrow BS_{B_1}  \rightarrow D_{B_1}$, $S_B \rightarrow C_1 \rightarrow BS_{B_1} \rightarrow BS_{B_0} \rightarrow FM_2 \rightarrow BS_{B_0} \rightarrow BS_{B_1} \rightarrow D_{B_1}$, and $S_B \rightarrow C_1 \rightarrow BS_{B_1} \rightarrow FM_3 \rightarrow BS_{B_1} \rightarrow D_{B_1}$.
      \end{enumerate}

\end{enumerate}

According the analysis of the optical paths, we obtain the detection probabilities of the detectors, which are listed in Table~\ref{tab:1} and~\ref{tab:2}. It can be seen that some detectors may receive two photons. In the analysis of the ability to attack, assume each detection equipment has the ability to measure the polarization and the number of the photons.
%

\begin{table}[!h]
\caption{The detection probability of each detector for the photon sent by Alice. $P_{D_0}$, $P_{D_1}$, $P_{D_2}$, $P_{D_{B_0}}$, and $P_{D_{B_1}}$ are he detection probabilities of detectors $D_0$, $D_1$, $D_2$, $D_{B_0}$ and $D_{B_1}$, respectively. $r_A$ and $t_A$ are the reflectivity and transmissivity of the beam splitter $BS_A$. $r_{B_0}$ and $t_{B_0}$ are the reflectivity and transmissivity of the beam splitter $BS_{B_0}$. $r_{B_1}$ and $t_{B_1}$ are the reflectivity and transmissivity of the beam splitter $BS_{B_1}$.}
\label{tab:1}       
\begin{tabular}[!h]{lll}
\hline\noalign{\smallskip}
~~ & $a_j^{(i)}\neq b_j^{(i)}$ & $a_j^{(i)}= b_j^{(i)}$ \\
\noalign{\smallskip}\hline\noalign{\smallskip}
$P_{D_0}$     & $r_A^2 +t_A^2 r_{B_0}^2$     & $r_A^2+t_A^2r_{B_0}^2$ \\
$P_{D_1}$     & $r_A t_A+ t_A r_{B_0}^2 r_A$ & $r_A t_A+ t_A r_{B_0}^2 r_A$ \\
$P_{D_2}$     & $t_A t_{B_0}$                & $t_A t_{B_0}$ \\
$P_{D_{B_0}}$ & $t_A r_{B_0} t_{B_0} t_{B_1}$ & $t_A r_{B_0} t_{B_0} t_{B_1}$ \\
$P_{D_{B_1}}$ & $t_A r_{B_0} t_{B_0} r_{B_1}$ & $t_A r_{B_0} t_{B_0} r_{B_1}$ \\
\noalign{\smallskip}\hline
\end{tabular}
\end{table}
\begin{table}[!h]
\caption{The detection probability of each detector for the photon sent by Bob.}
\label{tab:2}       
\begin{tabular}[!h]{lll}
\hline\noalign{\smallskip}
~~ & $a_j^{(i)}\neq b_j^{(i)}$ & $a_j^{(i)}= b_j^{(i)}$ \\
\noalign{\smallskip}\hline\noalign{\smallskip}
$P_{D_0}$     & $0$     & ${t_{B_1}t_{B_0}r_{B_0}t_A}$ \\
$P_{D_1}$     & $0$ & ${t_{B_1}t_{B_0}r_{B_0}r_A} $ \\
$P_{D_2}$     & $0$  & ${t_{B_1} r_{B_0}}$ \\
$P_{D_{B_0}}$ & $1$ & $t_{B_1}^2 t_{B_0}^2+{r_{B_1}^2}$ \\
$P_{D_{B_1}}$ & $0$ & $ {t_{B_1}t_{B_0}^2r_{B_1}}+{r_{B_1}t_{B_1}}$ \\
\noalign{\smallskip}\hline
\end{tabular}
\end{table}

\subsection{Probabilities related to the security ($P_A$ and $P_B$)}
By analyzing the response of detectors, Bob knows Alice's bit for sure with probability $P_B$. And the probability of Alice determining whether Bob knows he bit is  $P_A$.
In our protocol, Alice and Bob send a photon by their own single photon source. Therefore, each party at most detects two photons in the protocol. By classifying different number of the photons detected by Alice and Bob, analyze the probabilities $P_B$ and $P_A$. There are only three situations. (i) Both of the two photons go to Alice's site, while Bob detects none. (ii) One photon goes to Alice's site, while the other goes to Bob's. (iii) Both of the two photons go to Bob's site, while Alice detects none.
\begin{enumerate}
  \item Alice detects two photons. Two photons both get to Alice' site means that Bob detects no photon. If their bits are $a_j^{(i)}\neq b_j^{(i)}$, the photon sent by Bob must interfere and return to detector $D_{B_0}$. Therefore, when Alice detects two photons and Bob detects none, they know $a_j^{(i)}= b_j^{(i)}$. Below are the cases that Alice detects two photons.
      \begin{enumerate}
        \item Only one of Alice's detectors clicks. When detector $D_0$ detects two photon, the probability is  the probability that $D_0$ detects the photon sent by Alice times the probability that $D_0$ detects the photon sent by Bob in the case of $a_j^{(i)}= b_j^{(i)}$. According to Table~\ref{tab:1} and ~\ref{tab:2},
            \begin{equation}\label{2D0}
              P(two~photons~in~D_0)=\frac{1}{2}\left(r_A^2+t_A^2r_{B_0}^2\right)t_{B_1}t_{B_0}r_{B_0}t_A,
            \end{equation}
            where $1/2$ is the probability of the case $a_j^{(i)}= b_j^{(i)}$. Similarly, the probabilities of two photons detection of $D_1$ and $D_2$ are
             \begin{align}
               P(two~photons~in~D_1)&=\frac{1}{2}\left(r_A t_A+ t_A r_{B_0}^2 r_A\right){t_{B_1}t_{B_0}r_{B_0}r_A},\\
               P(two~photons~in~D_2)&=\frac{1}{2}t_A t_{B_0}{t_{B_1} r_{B_0}}.
             \end{align}
        \item Two of Alice's detectors click. When $a_j^{(i)}= b_j^{(i)}$, both detectors $D_0$ and $D_1$ may click. The photon sent by Alice and the photon sent by Bob are received by the detectors. And the detection probability of both $D_0$ and $D_1$ is
            \begin{equation}
            \begin{aligned}
              P(D_0,~D_1)=&\frac{1}{2}\big[\left(r_A^2+t_A^2r_{B_0}^2\right)t_{B_1}t_{B_0}r_{B_0}r_A\\
              &+\left(r_A t_A+ t_A r_{B_0}^2r_A\right)t_{B_1}t_{B_0}r_{B_0}t_A\big].
            \end{aligned}
            \end{equation}
            The detection probabilities of both $D_0$ and $D_2$, $D_1$ and $D_2$ are
            \begin{align}
               P(D_0,~D_2)&=\frac{1}{2}\left[\left(r_A^2+t_A^2r_{B_0}^2\right)t_{B_1} r_{B_0}+t_{B_1}t_{B_0}^2 r_{B_0}t_A^2\right],\\
               P(D_1,~D_2)&=\frac{1}{2}\left[\left(r_A t_A+ t_A r_{B_0}^2 r_A\right)t_{B_1} r_{B_0}+t_{B_1}t_{B_0}^2 r_{B_0}r_At_A\right].
            \end{align}
      \end{enumerate}

  \item Each party detects one photon. One photon goes to Alice's site while the other goes to Bob's. As Bob has only two detectors, either detector $D_{B_0}$ or detector $D_{B_1}$ receives a photon.
      \begin{enumerate}
        \item Detector $D_{B_0}$ clicks. When the bits chosen by them satisfy $a_j^{(i)}\neq b_j^{(i)}$, the photon sent by Bob will be interfere and make the detector response. When $a_j^{(i)}=b_j^{(i)}$, there is a possibility that photon sent by Alice or Bob goes to the detector. Whatever $a_j^{(i)}$ and $b_j^{(i)}$ are, detector $D_{B_0}$ may click. The polarization of the receiving photon is in accordance with the choice of Bob's bit, which makes Bob cannot distinguish what Alice sends.
        \item Detector $D_{B_1}$ clicks. In this case, detector $D_{B_0}$ receives no photon, which denotes $a_j^{(i)}=b_j^{(i)}$ and Bob knows what $a_j^{(i)}$ is. One of the photon goes to $D_{B_1}$ while the other one goes to Alice's site. The probability of this case is that
            \begin{align}\label{AandB}
              &P(D_{B_1},~Alice)\\\nonumber
              =&\frac{1}{2}\big[\big(r_A^2 +t_A^2 r_{B_0}^2+ r_A t_A+ t_A r_{B_0}^2 r_A+ t_A t_{B_0}\big)\big(t_{B_1}t_{B_0}^2r_{B_1}+r_{B_1}t_{B_1}\big)\\\nonumber
              &+\big(t_{B_1}t_{B_0}r_{B_0}t_A+ t_{B_1}t_{B_0}r_{B_0}r_A+ t_{B_1} r_{B_0}\big)t_A r_{B_0} t_{B_0} r_{B_1} \big]
            \end{align}
      \end{enumerate}
  \item Both of Bob's detectors click. For the two detectors, there are three cases as follows.
  \begin{enumerate}
    \item Detector $D_{B_0}$ receives two photons. Whatever $a_j^{(i)}$ and $b_j^{(i)}$ are, the two photon have possibility to get to $D_{B_0}$. When Alice's and Bob's bits satisfy $a_j^{(i)}\neq b_j^{(i)}$, the polarizations of the two photon are different. Otherwise, the polarizations are the same. Bob could know Alice's bit by the polarizations. The probability of confirming Alice's bit in this case is
        \begin{equation}
          P(two~photons~in~D_{B_0})=\frac{1}{2}t_A r_{B_0} t_{B_0} t_{B_1}+\frac{1}{2}t_A r_{B_0} t_{B_0} t_{B_1}\left(t_{B_1}^2 t_{B_0}^2+r_{B_1}^2\right)
        \end{equation}
    \item Detector $D_{B_1}$ receives two photons. $a_j^{(i)}\neq b_j^{(i)}$ leads at least one photon go to detector $D_{B_0}$. Therefore, Bob knows that $a_j^{(i)}=b_j^{(i)}$, when $D_{B_1}$ receives two photons. The probability of two photons in $D_{B_1}$ is
        \begin{equation}
          P(two~photons~in~D_{B_1})=\frac{1}{2}t_A r_{B_0} t_{B_0} r_{B_1}\left(t_{B_1}t_{B_0}^2r_{B_1}+r_{B_1}t_{B_1}\right)
        \end{equation}
    \item Both detectors $D_{B_0}$ and $D_{B_1}$ receive one photon. Bob confirms Alice's bit by the polarizations of the different photons. With the different choices of $a_j^{(i)}$ and $b_j^{(i)}$, the probabilities are
        \begin{equation}
        \begin{aligned}
          P(D_{B_0},~D_{B_1})=&\frac{1}{2}t_A r_{B_0} t_{B_0} r_{B_1}+
          \frac{1}{2}\big[t_A r_{B_0} t_{B_0} t_{B_1}\big(t_{B_1}t_{B_0}^2r_{B_1}\\ &+r_{B_1}t_{B_1}\big)
          +\big(t_{B_1}^2 t_{B_0}^2+ r_{B_1}^2\big)t_A r_{B_0} t_{B_0} r_{B_1}\big]  \label{2Bob}
        \end{aligned}
        \end{equation}
  \end{enumerate}
\end{enumerate}

Through the analysis above, it can be seen that when both photons received by one of the party, Alice knows that Bob confirms the value of her bit. The probability $P_A$ is the sum of Eq.~(\ref{2D0}) to Eq.~(\ref{2Bob}) except Eq.~(\ref{AandB}). For each beam splitter, the sum of the reflectivity and the transmissivity is $1$, i.e. $r_A+t_A=1$, $r_{B_0}+t_{B_0}=1$ and $r_{B_1}+t_{B_1}=1$. The reduction result of the probability $P_A$ is
\begin{equation}
\begin{aligned}\label{pA}
  P_A=&\frac{1}{2}\big[t_At_{B_0}+2t_At_{B_0}t_{B_1}-5t_At_{B_0}t_{B_1}^2+2t_At_{B_0}t_{B_1}^3-t_At_{B_0}^2-2t_At_{B_0}^2t_{B_1}\\
  &+5t_At_{B_0}^2t_{B_1}^2-2t_At_{B_0}^2t_{B_1}^3+3t_At_{B_0}^3t_{B_1}-3t_At_{B_0}^3t_{B_1}^2+2t_At_{B_0}^3t_{B_1}^3\\
  &-3t_At_{B_0}^4t_{B_1}+3t_At_{B_0}^4t_{B_1}^2-2t_At_{B_0}^4t_{B_1}^3-t_{B_0}^2t_{B_1}+t_{B_1}\big].
\end{aligned}
\end{equation}
 The probability $P_B$ is the sum of Eq.~(\ref{2D0}) to Eq.~(\ref{2Bob}). And the reduction result is
\begin{equation}
\begin{aligned}\label{pB}
  P_B=&\frac{1}{2}\big[t_At_{B_0}+2t_At_{B_0}t_{B_1}-5t_At_{B_0}t_{B_1}^2+2t_At_{B_0}t_{B_1}^3-t_At_{B_0}^2-2t_At_{B_0}^2t_{B_1}\\
  &+5t_At_{B_0}^2t_{B_1}^2-2t_At_{B_0}^2t_{B_1}^3+t_At_{B_0}^3t_{B_1}-t_At_{B_0}^3t_{B_1}^2+2t_At_{B_0}^3t_{B_1}^3\\
  &-t_At_{B_0}^4t_{B_1}+t_At_{B_0}^4t_{B_1}^2-2t_At_{B_0}^4t_{B_1}^3-t_{B_0}^2t_{B_1}^2+2t_{B_1}-t_{B_1}^2\big]
\end{aligned}
\end{equation}

\section{Appropriate security parameters}
\subsection{Security against malicious choice of optimal beam splitters}\label{m,n}
It can be seen that the probabilities $P_A$ and $P_B$ are determined by the parameters of three beam splitters. According to Fig.~\ref{fig:QBC}, the beam splitter $BS_A$ is in Alice's own site, $BS_{B_0}$ and $BS_{B_1}$ both belong to Bob. That is, a malicious party may not use the half transparent and half reflecting mirror. Instead, he (she) uses the devices in favor of his (her) attack.

We first consider the situation where both parties are honest. In this situation, all of the beam splitters are half transparent and half reflecting mirrors, i.e. $t_A=t_{B_0}=t_{B_1}=1/2$. According to Eq.~(\ref{pA}) and (\ref{pA}), the probabilities are
\begin{equation}
  P_A=\frac{17}{64},~~~P_B=\frac{53}{128}.
\end{equation}
The probability that Alice alters one bit of a sequence without detection is
\begin{equation}
  P(Aalter)=\frac{1-P_B}{1-P_A}=\frac{75}{94}.
\end{equation}
To satisfy the security threshold $\alpha$ and $\beta$, according to Eq.~(\ref{threshold of m}) and (\ref{threshold of n}) the range of $m$ and $n$ are
\begin{align}
   &m>\frac{\log\alpha}{\log P(Aalter)}=\frac{\log\alpha}{\log ({75}/{94})},\label{m}\\
   &n>\frac{\log[1-(1-2\beta)^{1/m}]}{\log P_B}=\frac{\log[1-(1-2\beta)^{1/m}]}{\log ({53}/{128})},\label{n}
\end{align}
respectively. 

However, Alice and Bob may try to attack the protocol. The binding security is determined by the probability $P(Aalter)$ and the security parameter $m$. The concealing security is determined by the probability $P_B$ and the two security parameters $m$ and $n$. The larger the security parameters are, the more secure the protocol is. We first analyze Alice's optimal strategy of the maximum $P'(Aalter)$.  Then analyze Bob's optimal strategy of the maximum $P''_B$. To protect both binding and concealing security, there are appropriate value ranges of $m$ and $n$.

A malicious Alice attacks the binding of QBC protocol, who can optimize the $BS_A$ to make a larger $P'(Aalter)$. In this case, Bob sets up his beam splitters as half transparent and half reflecting mirrors, $t_{B_0}=t_{B_1}=1/2$. The probabilities $P'_A$ and $P'_B$ are
\begin{equation}
  P'_A=\frac{5}{32}t_A+\frac{3}{16},~~~P'_B=\frac{9}{64}t_A+\frac{11}{32}.
\end{equation}
The probability that Alice alters one bit of a sequence without detection is
\begin{equation}
  P'(Aalter)=\frac{1-P_B}{1-P_A}=\frac{42-9t_A}{52-10t_A}.
\end{equation}
Alice can set up the beam splitter $BS_A$ as $t_A=0$ to get the maximum $P'(Alater)$. When $t_A=0$, $P'(Aalter)=\frac{21}{26}$, $P'_B=\frac{11}{32}$.
Even if Alice changes the beam splitter, the binding security can still be achieved by choosing appropriate secure parameters. To satisfy the security threshold $\alpha$ and $\beta$, $m$ and $n$ should satisfy
\begin{align}
   &m>\frac{\log\alpha}{\log P'(Aalter)}=\frac{\log\alpha}{\log ({21}/{26})},\label{m'}\\
   &n>\frac{\log[1-(1-2\beta)^{1/m'}]}{\log P'_B}=\frac{\log[1-(1-2\beta)^{1/m'}]}{\log ({11}/{32})},
\end{align}
respectively. 
It can be seen that the threshold of Eq.~(\ref{m'}) is lager than that of Eq.~(\ref{m}) considered Alice's attack. To ensure the binding security of the protocol, the range of $m$ in Eq.~(\ref{m'}) is appropriate.

A malicious Bob attacks the concealing of QBC protocol, who can optimize the beam splitters $BS_{B_0}$ and $BS_{B_1}$ to obtain a larger $P''_B$. In this case, Alice's beam splitter is a half transparent and half reflecting mirror, $t_A=1/2$. For each qubit, the probability that Bob knows for sure what Alice sends is
\begin{equation}
  \begin{aligned}
    P''_B=&\frac{1}{2}\big(t_{B_0}/2+t_{B_0}t_{B_1}-5t_{B_0}t_{B_1}^2/2+t_{B_0}t_{B_1}^3-t_{B_0}^2/2-t_{B_0}^2t_{B_1}+3t_{B_0}^2t_{B_1}^2/2\\
    &-t_{B_0}^2t_{B_1}^3+t_{B_0}^3t_{B_1}/2-t_{B_0}^3t_{B_1}^2/2+t_{B_0}^3t_{B_1}^3-t_{B_0}^4t_{B_1}/2+t_{B_0}^4t_{B_1}^2/2\\
    &-t_{B_0}^4t_{B_1}^3+2t_{B_1}-t_{B_1}^2\big)
  \end{aligned}
\end{equation}

\begin{figure}[!h]
\centering
\includegraphics[width=0.5\textwidth]{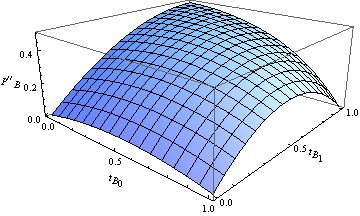}
\caption{When $t_A=1/2$, the probability $P''_B$ that Bob knows for sure what Alice sends varies with different transmissivities of the beam splitters $BS_{B_0}$ and $BS_{B_1}.$}
\label{fig:PB}       
\end{figure}

Fig.~{\ref{fig:PB}} shows the probability of $P''_B$. When $t_{B_0}=0$, $t_{B_1}=1$, the maximum of $P''_B$ is $1/2$. To protect the QBC protocol from Alice's attack, Bob agrees the security parameter $m$ according to Eq.~(\ref{m'}). Given a security threshold $\beta$, $n$ should satisfy
\begin{equation}
  n>\frac{\log[1-(1-2\beta)^{1/m}]}{\log P''_B}=-\log[1-(1-2\beta)^{1/m}].\label{n''}
\end{equation}
 It can be seen that the threshold of Eq.~(\ref{n''}) is lager than that of of Eq.~(\ref{n}), which is considered Bob's attack. To ensure the concealing security of the protocol, the range of $n$ in Eq.~(\ref{n''}) is secure. Therefore, the range of appropriate security parameters should satisfy Eq.~(\ref{m'}) and Eq.~(\ref{n''}).

When set up the specific security threshold, for example $\alpha=10^{-6}$ and $\beta=10^{-6}$, the security parameter $m$ can be computed as $m>64.7$ according to Eq.~(\ref{m'}). Assumed $m=65$, the range of security parameter $n$ is $n>24.9$, which can be set up as $n=25$. Therefore, when $m=65$, $n=25$, no matter what kind of beam splitter chosen by a malicious party the binding and concealing satisfy the presupposed security requirement.

\subsection{Security against malicious party sending advantageous states}
To strictly execute the protocol, Alice and Bob should send single-photon state. A malicious party may try to send no photon or multi-photon states while the other party is honest to maintain the protocol. The case that one of the parties sends multi-photon states leads the detectors to detect multi-photon states. As long as the other party's detectors response, she (he) discovers the attack. For one of the parties sends no photon, the analysis is as follows:

If Bob sends no photon, it means that Bob does not choose his bit $b_j^{(i)}$. Only the photon sent by Alice is transmitted in the optical system. The detection probability of each detector when Bob sends no photon is listed in Table~\ref{tab:tb1}. Bob can measure the polarization of Alice's photon by his detector. The probability that Bob confirms Alice's bit is $P_B(Bob~none)$,
\begin{equation}
  P_B(Bob~none)=P_{D_{B_0}}(Bnone)+P_{D_{B_1}}(Bnone)=t_Ar_{B_0}t_{B_0}\leq t_A.
\end{equation}
Considered there is only one malicious party, Alice strictly executes the protocol, i.e. $t_A=1/2$. In this case, $P_B(Bnone)\leq P''_B$, the security parameters determined in Subsection~\ref{m,n} are still appropriate.

\begin{table}[!h]
\caption{ The detection probability of each detector when Bob sends no photon.}
\label{tab:tb1}       
\begin{tabular}[!h]{lll}
 \hline\noalign{\smallskip}
 ~~ & Bob sends no photon &~\\
 \noalign{\smallskip}\hline\noalign{\smallskip}
$P_{D_0}(Bnone)$     & $r_A^2 +t_A^2r_{B_0}^2$    \\
$P_{D_1}(Bnone)$     & $r_A t_A+t_Ar_{B_0}^2r_A$ \\
$P_{D_2}(Bnone)$     & $t_At_{B_0}$               \\
$P_{D_{B_0}}(Bnone)$ & $t_Ar_{B_0}t_{B_0}t_{B_1}$ \\
$P_{D_{B_1}}(Bnone)$ & $t_Ar_{B_0}t_{B_0}r_{B_1}$  \\
\noalign{\smallskip}\hline
\end{tabular}
\end{table}

\begin{table}[!h]
\caption{The detection probabilities of each detector when Alice sends no photon.}
\label{tab:tb2}       
\begin{tabular}[!h]{lll}
\hline\noalign{\smallskip}
 ~&$a_j^{(i)}\neq b_j^{(i)}$ & $a_j^{(i)}= b_j^{(i)}$ \\
 \noalign{\smallskip}\hline\noalign{\smallskip}
$P_{D_0}(Anone)$     &~~~~0 & ${t_{B_1}t_{B_0}r_{B_0}t_A}$ \\
$P_{D_1}(Anone)$     &~~~~0 & ${t_{B_1}t_{B_0}r_{B_0}r_A} $ \\
$P_{D_2}(Anone)$     &~~~~0 & ${t_{B_1}r_{B_0}}$ \\
$P_{D_{B_0}}(Anone)$ &~~~~1 & ${t_{B_1}^2 t_{B_0}^2}+ {r_{B_1}^2}$ \\
$P_{D_{B_1}}(Anone)$ &~~~~0 & $t_Bt_{B_0}^2r_{B_1}+r_Bt_{B_1}$ \\
\noalign{\smallskip}\hline
\end{tabular}
\end{table}

When Alice sends no photon, Bob strictly executes the protocol, i.e. $t_{B_0}=t_{B_1}=1/2$. Alice still chooses her bit $a_j^{(i)}$ because of the existence of detector $D_2$. The detection probabilities of each detector when Alice sends no photon are listed in Tabel~\ref{tab:tb2}. Bob treats the cases that he detects two photons, $D_{B_1}$ clicks, and detects no photon as $a_j^{(i)}= b_j^{(i)}$. When Alice sends no photon, Bob can never detect two photons. And the probability he confirms $a_j^{(i)}= b_j^{(i)}$ is $P_B(Anone)$,
\begin{equation}
\begin{aligned}
  P_B(Anone)&=\frac{1}{2}\big(P_{D_0}(Anone)+P_{D_1}(Anone)+P_{D_2}(Anone)+P_{D_{B_1}}(Anone)\big)\\
  &=11/32.
\end{aligned}
\end{equation}
In this case, only when Alice's detectors respond, she ensures Bob knows her bit, the probability of which is
\begin{equation}
  P_B(Anone)=\frac{1}{2}\big(P_{D_0}(Anone)+P_{D_1}(Anone)+P_{D_2}(Anone)\big)=3/16.
\end{equation}
Therefore, the probability Alice alters one bit in a sequence without detection is
\begin{equation}
  P(Aalter~Anone)=\frac{1-P_B(Anone)}{1-P_A(Anone)}=21/26=P'(Aalter).
\end{equation}
The security parameters determined in Subsection~\ref{m,n} can still keep it secure.

\section{The relation between our QBC and no-go theorem}
The framework of  no-go theorem is described as follows.
\begin{enumerate}
  \item When Alice commits $b$, she prepares
\begin{equation}\label{no-go state}
|\Phi_b\rangle=\sum_{i} \alpha_i^{(b)}\big|e_i^{(b)}\big\rangle_A\otimes \big|\phi_{i}^{(b)}\big\rangle_B,
\end{equation}
where $\big\langle e_i^{(b)}\big|e_j^{(b)}\big\rangle_A=\delta_{ij}$ while $\big|\phi_{i}^{(b)}\big\rangle_B$'s are not necessarily orthogonal to each other.
She sends the second register to Bob as a piece of evidence.
\item To ensure the concealing of the QBC protocol, the density matrices describing the second register are approximative. i.e.,
\begin{equation}
\label{concealing}
  Tr_A|0\rangle\langle0|\equiv\rho_0^B\simeq \rho_1^B\equiv Tr_A|1\rangle\langle1|.
\end{equation}
\item When Eq.~(\ref{concealing}) is satisfied, Alice can apply a local unitary transformation $U_A$ to rotate $|0\rangle$ to $|1\rangle$ without detection.
\end{enumerate}

We will demonstrate that our protocol is not fit the framework of no-go theorem for three situations. (i) Alice sends the states according to the protocol; (ii) Alice prepares $2n$-qubits entangled states and sends part of them; (iii) Alice prepares $n$-qubits entangled states and sends all of them.

\subsection{The case that Alice sends the states honestly}
In the commit phase of our protocol, Alice and Bob send $m$ qubit strings with the length of $n$, respectively. It is equivalent to the case that execute the $n$ qubits protocol $m$ times. In the following analysis, we only consider one qubit string with the length of $n$ firstly.
The ith sequence sent by Alice contains $n$ qubits $\big|\Psi_A\big\rangle\equiv\big|\Psi_{a_1}\big\rangle\otimes
\big|\Psi_{a_2}\big\rangle\otimes...\otimes\big|\Psi_{a_n}\big\rangle$, while the sequence sent by Bob is $\big|\Psi_B\big\rangle\equiv\big|\Psi_{b_1}\big\rangle\otimes
\big|\Psi_{b_2}\big\rangle\otimes...\otimes\big|\Psi_{b_n}\big\rangle$. The initial state of the system is $\big|\Psi_A\big\rangle\otimes\big|\Psi_B\big\rangle$, which contains $2n$ qubits. After the commit phase of the protocol, some of the $2n$ qubits are detected by Alice, others are detected by Bob. The final state in Alice's hands is
\begin{equation}
  \big|\Psi'_A\big\rangle\equiv\big|\Psi_{a_{\mu_1}}\big\rangle\otimes
\big|\Psi_{a_{\mu_2}}\big\rangle\otimes...\otimes\big|\Psi_{a_{\mu_l}}\big\rangle \big|\Psi_{b_{\nu_1}}\big\rangle\otimes\big|\Psi_{b_{\nu_2}}\big\rangle\otimes... \otimes\big|\Psi_{b_{\nu_k}}\big\rangle,
\end{equation}
where $\mu_1, \mu_2,...,\mu_l$ is the label of qubits sent by Alice and detected by Alice, $\nu_1, \nu_2,...,\nu_k$ is the label of qubits sent by Bob and detected by Alice.
The final state in Bob's hands is
\begin{equation}
  \big|\Psi'_B\big\rangle\equiv\big|\Psi_{a_{\mu_{l+1}}}\big\rangle\otimes
\big|\Psi_{a_{\mu_{l+2}}}\big\rangle\otimes...\otimes\big|\Psi_{a_{\mu_n}}\big\rangle \big|\Psi_{b_{\nu_{k+1}}}\big\rangle\otimes\big|\Psi_{b_{\nu_{k+2}}}\big\rangle\otimes... \otimes\big|\Psi_{b_{\nu_n}}\big\rangle,
\end{equation}
where $\mu_{l+1}, \mu_{l+2},...,\mu_n$ is the label of qubits sent by Alice and detected by Bob, $\nu_{k+1}, \nu_{k+2},...,\nu_n$ is the label of qubits sent by Bob and detected by Bob.

In our protocol, the qubits in Alice's hand and the qubits in Bob's hand are independent and different, which is different from Eq.~(\ref{no-go state}). No matter what operation Alice applies on the registes in her's hand, it cannot change the states in Bob's hand. To attack the binding security without detection, the only way is to change the qubit in her own site. However, Bob may knows Alice's bit even if he does not detect the photons, which is classical security analyzed in Section 5.  Our QBC protocol does not fit the framework of no-go theorem.

\subsection{The case that Alice prepares the entangled states and sends part of them}
If Alice prepares the entangled state as Eq.~(\ref{no-go state}), i.e.
\begin{equation}
  |\Phi_b\rangle_A=\sum_{\bigoplus_j a_j=b} \alpha_{a^{(i)}}^{(b)}\big|e_{a^{(i)}}^{(b)}\big\rangle_A\otimes \big|\Psi_{a^{(i)}}^{(b)}\big\rangle_{patha},
\end{equation}
where $\big|e_{a^{(i)}}^{(b)}\big\rangle_A$ is an orthogonal basis of system A. She keeps the register $A$ and sends the second register through the path $a$ in fig.~\ref{fig:QBC}.
Bob sends the single-photon states honestly as $|\Psi_{b_1}\rangle\otimes|\Psi_{b_2}\rangle\otimes...\otimes|\Psi_{b_n}\rangle$.
The initial state of the whole system is
\begin{equation}
 \begin{aligned}
  |\Phi_b\rangle_{AB}=\sum_{\bigoplus_j a_j=b} \alpha_{a^{(i)}}^{(b)}\big|e_{a^{(i)}}^{(b)}\big\rangle_A\otimes \big|\Psi_{a^{(i)}}^{(b)}\big\rangle_{patha} \otimes|\Psi_{b_1}\Psi_{b_2}...\Psi_{b_j}...\Psi_{b_n}\rangle_B
 \end{aligned}
\end{equation}

Bob's detection leads the state to collapse to the other state. The qubits which reach to detectors $D_{B_0}$ and $D_{B_1}$ are probably sent by Alice or Bob. If Alice measures the qubits which goes to her detectors, all of the $2n$ qubits are detected and the entanglement between the system A and the other registers will be completely destroyed, which is not satisfied the condition of no-go theorem. Therefore, she does not measure the qubits which goes to her detectors in the beginning. According to the no-go theorem, if the density matrices $\rho_0^B\simeq \rho_1^B$, Alice can apply a local unitary transformation to alter the commit.

There are two possible strategies for Alice to apply no-go theorem type attack. One is to treat the unsent register A as the register that she would apply the local unitary transformation on. The difference between the states of $|\Phi_0\rangle_{AB}$ and $|\Phi_1\rangle_{AB}$ is that at least one qubit $|\Psi_{a_j}\rangle$ are different. Because the states $|\Psi_0\rangle=|H\rangle$ and $|\Psi_1\rangle=|V\rangle$ are orthogonal, the states $|\Phi_0\rangle_{AB}$ and $|\Phi_1\rangle_{AB}$ are orthogonal. Therefore, the density matrices satisfies $\rho_0^{aB}\perp \rho_1^{aB}$, which leads to the inability of the no-go theorem. The other strategy is to apply the local unitary transformation on the register A and the qubits which are sent by Alice and finally come back to Alice. If the qubits sent by Alice and detected by Bob are the same when $b=0$ and $b=1$, the density matrices describing the registers in Bob's hand are the same, i.e. $\rho_0^{B}= \rho_1^{aB}$, which makes it possible for no-go theorem type attack. However, for keeping the entanglement, Alice cannot measure the qubits in her hand before the attack. It leads that she has no idea about the exactly states she would apply the local unitary transformation. Therefore, she cannot apply the no-go theorem type attack in this way.

\subsection{The case that Alice prepares the entangled states and sends all of them}
Alice may try to send another entangled state containing $n$ qubits. The aim of the operation is to make the qubits sent by Alice finally in Alice's hands are entangled with those in Bob's hands. Assume that the state sent by Alice is
\begin{equation}
  \frac{1}{\sqrt{2^n}}\sum_{\bigoplus_j a_j=b} |\Psi_{a_1}\Psi_{a_2}...\Psi_{a_j}...\Psi_{a_n}\rangle.
\end{equation}
The initial state of the whole system is
\begin{equation}
   \frac{1}{\sqrt{2^n}}\sum_{\bigoplus_j a_j=b} |\Psi_{a_1}\Psi_{a_2}...\Psi_{a_j}...\Psi_{a_n}\rangle \otimes|\Psi_{b_1}\Psi_{b_2}...\Psi_{b_j}...\Psi_{b_n}\rangle.
\end{equation}
After executing the protocol, some of the qubits go to Alice's site and the others go to Bob's. Bob measures the qubits honestly while Alice does not measure temporarily. After Bob's measurement, the state becomes
\begin{equation}
\begin{aligned}\label{attack3}
  |\Psi_{a_{\mu_{l+1}}}\rangle\otimes
|\Psi_{a_{\mu_{l+2}}}\rangle\otimes...\otimes|\Psi_{a_{\mu_n}}\rangle |\Psi_{b_{\nu_{k+1}}}\rangle\otimes|\Psi_{b_{\nu_{k+2}}}\rangle\otimes... \otimes|\Psi_{b_{\nu_n}}\rangle \\ \otimes\sum_{a_{\mu_j},b_{\nu_j}}|\Psi_{a_{\mu_1}}\rangle\otimes
|\Psi_{a_{\mu_2}}\rangle\otimes...\otimes|\Psi_{a_{\mu_l}}\rangle |\Psi_{b_{\nu_1}}\rangle\otimes|\Psi_{b_{\nu_2}}\rangle\otimes... \otimes|\Psi_{b_{\nu_k}}\rangle,
\end{aligned}
\end{equation}
where $\mu_1, \mu_2,...,\mu_l$ is the label of qubits sent by Alice and go to Alice, $\nu_1, \nu_2,...,\nu_k$ is the label of qubits sent by Bob and go to Alice. $\mu_{l+1}, \mu_{l+2},...,\mu_n$ is the label of qubits sent by Alice and detected by Bob, $\nu_{k+1}, \nu_{k+2},...,\nu_n$ is the label of qubits sent by Bob and detected by Bob. According to no-go theorem, Alice can alter her commit bit by the local unitary transformation. The precondition of no-go theorem is that Alice knows what the entangled state is. In $\sum_{a_{\mu_j},b_{\nu_j}}|\Psi_{a_{\mu_1}}\rangle\otimes
|\Psi_{a_{\mu_2}}\rangle\otimes...\otimes|\Psi_{a_{\mu_l}}\rangle |\Psi_{b_{\nu_1}}\rangle\otimes|\Psi_{b_{\nu_2}}\rangle\otimes... \otimes|\Psi_{b_{\nu_k}}\rangle$ of Eq.~(\ref{attack3}), Alice has no idea about the qubits sent by Bob. Therefore, she cannot find the local unitary transformation to attack. Even she found the unitary transformation and altered the commit bit, she cannot reveal the correct states sent by Bob and detected by Bob, which makes the attack fails.

\section{Discussion}
The protocol we present in this paper is an ideal scheme without considering the error and loss in practice. Once there were error and loss in the quantum channel, the detection would be different and influenced the security of the scheme. Moreover, how to hold the stability of the Michelson-type interferometer for long distance is the question. It's worth considering to improve our scheme by Sagnac interferometer. In Section 3, we propose a framework for QBC without considering no-go theorem. How to construct universal QBC framework which can resist no-go theorem type attack like the scheme we proposed in this paper is worth to discuss. We would like to leave these questions open for future researches.

\section{Conclusion}
In this paper, we present a new QBC protocol based on counterfactual quantum cryptography. We analyze the security of this protocol, find that it can resist the attack presented by QBC's no-go theorem. There are two essential reasons. (i) The states for different commit bit are orthogonal. (ii) Alice cannot distinguish the qubits that reach to her detectors sent by whom. Then she does not know which registers she could apply the unitary transformation $U_A$ of no-go theorem, which probably gives a new way of constructing QBC protocols.

\end{document}